\documentclass[preprint,showkeys,aps,prb]{revtex4}
\usepackage[dvips]{graphicx}

\begin{document}

\title{Why Instantaneous Values of the ``Covariant'' Lyapunov Exponents Depend upon
the Chosen State-Space Scale}

\author{
Wm. G. Hoover and Carol G. Hoover               \\
Ruby Valley Research Institute                  \\
Highway Contract 60, Box 601                    \\
Ruby Valley, Nevada 89833                       \\
}

\date{\today}


\vspace{0.1cm}

\begin{abstract}

We explore a simple example of a chaotic thermostated harmonic-oscillator system which exhibits qualitatively
different local Lyapunov exponents for simple scale-model constant-volume transformations of its coordinate
$q$ and momentum $p$ : $\{ \ q,p \ \} \rightarrow \{ \ (Q/s),(sP) \ \}$ .  The time-dependent thermostat
variable $\zeta(t)$ is unchanged by such scaling.  The original $(qp\zeta)$ motion and the scale-model
$(QP\zeta)$ version of the motion are physically identical.  But both the local Gram-Schmidt Lyapunov exponents
and the related local ``covariant'' exponents change with the change of scale.  Thus this model furnishes a
clearcut chaotic time-reversible example showing how and why both the local Lyapunov exponents and covariant
exponents vary with the scale factor $s$.

\end{abstract}

\maketitle

\section{Local/Global Gram-Schmidt/Covariant Vectors/Exponents}

The popularity of the time-dependent (or ``instantaneous'', or ``local'') covariant Lyapunov vectors
and their associated exponents as descriptions of chaotic motion seems to us to be linked to a (false)
impression extracted from the literature.  Some of the literature implies that these descriptors have
a special significance independent of such details as the coordinate system used to describe them.  A
selected literature, some of it quite clear, can be found in References 1-7.  If the chosen coordinate
system were {\it really} insignificant it would be hard to understand a simple, but nonchaotic,
counterexample : the one-dimensional harmonic oscillator, which exhibits a strong dependence of its
largest local Lyapunov exponent $\lambda_1(t)$ on the chosen Cartesian coordinate system\cite{b1,b8,b9}.

We remind the reader that this local instantaneous Lyapunov exponent $\lambda_1(t)$ (the largest of them
when time averaged) measures the local rate of divergence of two nearby trajectories.  Think of them as
a reference trajectory and a satellite trajectory, with the satellite constrained to remain near the
reference.  It is unnecessary to consider exponents beyond the first to understand why it is that the
local Lyapunov exponents, covariant or not, are in fact {\it not} scale-independent and do indeed depend
upon the chosen coordinate system or set of measurement units. The oft-repeated statement that the local 
covariant exponents are ``norm-independent'' should not be misunderstood (as we did) to mean that the
exponents are independent of a scale factor, as in a change of units from cgs to MKS.

\begin{figure}
\vspace{1 cm}
\includegraphics[height=7cm,width=6cm,angle=-90]{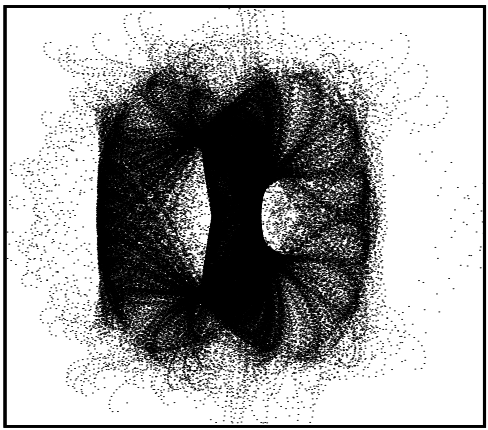}
\caption{
Chaotic attractor $\zeta(q)$ projection for $\epsilon = 0.20$ with fourth-order Runge-Kutta
timestep $dt = 0.001$ using $ 200,000$ equally-spaced points from the last half of a $40,000,000$
timestep simulation.  The abscissa and ordinate scales range from $-4.0$ to $+4.0$ .
}
\end{figure}

Here we focus on a simple chaotic continuous-flow example\cite{b10}, the thermostated three-dimensional flow of
a harmonic oscillator with coordinate $q$ , momentum $p$ , and friction coefficient $\zeta(t)$ in the
unscaled $(q,p,\zeta)$ phase space :
$$
\dot q = p \ ; \ \dot p = -q - \zeta p \ ; \ \dot \zeta = p^2 - T(q) \ ; \ 
T(q) = 1 + \epsilon \tanh (q) \ .
$$
The variation of temperature with coordinate $T(q)$ makes possible dissipation, and phase-volume
shrinkage, $\dot \otimes < 0$ , onto a torus, or a strange attractor with fractional dimensionality,
or a one-dimensional limit cycle.  For the evolution of this model see References 11-14.

For simplicity's sake the oscillator mass and force constant, as well as Boltzmann's constant, are all
chosen equal to unity here.  For $\epsilon = 0.20$ and with initial values $( \ q = 0, \ p = 5, \
\zeta = 0 \ )$ the motion generates a chaotic strange attractor, with two {\it time-averaged} nonzero
Lyapunov exponents $\lambda_1 \simeq + 0.01 \ ; \ \lambda_3 \simeq -0.01$ and with a time-averaged rate
of phase-volume contraction imposed by the friction coefficient $\zeta$ ,
$$
\langle \ \dot \otimes/\otimes \ \rangle = \langle \ (\partial \dot q/\partial q) +
(\partial \dot p/\partial p) + (\partial \dot \zeta /\partial \zeta) \ \rangle =
0 - \langle \ \zeta \ \rangle + 0 =  \lambda_1 + \lambda_2 + \lambda_3   \simeq - 0.0003 \ .
$$

Regular, limit-cycle, and chaotic solutions can all be found by following the related
work carried out in Reference 10.  These solutions' details depend upon the initial conditions
as well as the value of the maximum temperature gradient $\epsilon \equiv (dT/dq)_{q=0}$ .

Figures 1 and 2 show both a typical chaotic strange attractor ( positive $\lambda_1$ , generated
with $\epsilon = 0.20$ ) and an unusually elaborate limit cycle ( zero $\lambda_1$, generated with
$\epsilon = 0.37$ ).  The time required for the appearance of such limit cycles can be hundreds of
millions, or even billions, of timesteps.  Although fourth-order Runge-Kutta timesteps ranging from
0.0005 to 0.05 produce such a cycle a careful look at Figure 2 reveals a disconcerting dependence of cycle
topology on the time step (!).

\begin{figure}
\vspace{1 cm}
\includegraphics[height=7cm,width=6cm,angle=-90]{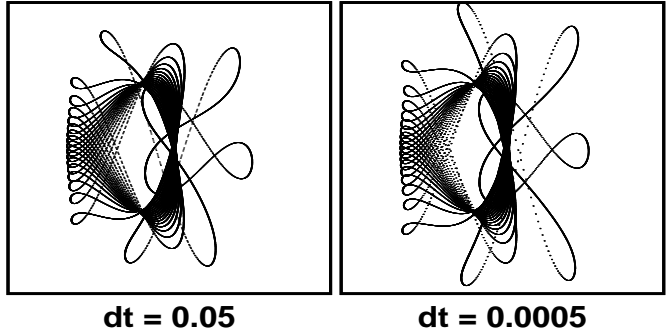}
\caption{
Limit cycles' $\zeta(q)$ projections for $\epsilon = 0.37$ using timesteps of $0.05$ and $0.0005$ .
The abscissa and ordinate range from $-4.0$ to $+4.0$ .
}
\end{figure}

The ``local'' time-dependent value of the largest Lyapunov exponent $\lambda_1(t)$ describes the rate
at which two nearby $(q,p,\zeta)$ trajectories tend to separate :
$$
\lambda_1(t) \equiv (d\ln r/dt) \ ; \ r \equiv \sqrt{\delta q^2 + \delta p^2 + \delta \zeta ^2}
\simeq e^{+\lambda_1t} \ .
$$

In the simple Gram-Schmidt picture ( and unlike the covariant picture with its nonorthogonal, but
still normalized  vectors ) adding in the {\it second} Lyapunov exponent $\lambda_2$ gives the rate
of divergence of the {\it area} defined by three nearby trajectories ( the reference and two 
satellites ), $\propto \exp{[ \ + \lambda_1t + \lambda_2t \ ]}$ .  The third Gram-Schmidt exponent
is needed to describe the divergence ( or shrinkage ) rate associated with the {\it volume} associated
with four nearby trajectories , $\propto \exp{[ \ + \lambda_1t + \lambda_2t + \lambda_3t \ ]}$ .
In these three Gram-Schmidt definitions the time $t$ is understood to be sufficiently long for
convergence of the exponents.

Typically, these time-averaged exponents don't depend on the coordinate system used to describe the
system because the divergence is {\it exponential}, and so depends only on the units of time, not
those of space or momentum.  Two identical chaotic systems, one described with MKS units and the
other with cgs units exhibit the {\it same} ( time-averaged ) rates of divergence even though the
mass and length scales differ.  It is also possible, usual, and useful to define ``local'' or
``instantaneous'' Lyapunov exponents by following two or more constrained trajectories and measuring
their tendencies to separate or approach each other as a function of the time of
measurement\cite{b1,b2,b3,b4,b5,b6,b7,b8,b9,b15}.  The MKS and cgs values of these local exponents differ.
The Gram-Schmidt Lyapunov exponents are simply the time averages of these instantaneous values :
$$
\lambda_1 = \langle \ \lambda_1(t) \ \rangle \ ; \
\lambda_2 = \langle \ \lambda_2(t) \ \rangle \ ; \
\lambda_3 = \langle \ \lambda_3(t) \ \rangle \ \dots \ .
$$

In typical situations, time-reversible and phase-volume-conserving Hamiltonian systems have ``paired''
Gram-Schmidt exponents, with the instantaneous identities :
$$
\lambda _1(t) + \lambda_N(t) = \lambda _2(t) + \lambda_{N-1}(t) = \dots \equiv 0 \ .
$$
But in exceptional cases ( like the collision of two many-body chunks of solid )\cite{b15} this pairing
can be violated.  Because the pairing reflects the time-reversibility of the Hamiltonian equations of
motion this lost symmetry is simply a symptom that the ``past'' can be sufficiently different to the
future. But because the time-averaged Hamiltonian exponents often exist in $\pm \lambda(t)$ pairs, the
largest exponent in either time direction is typically equal to (the negative of) the smallest exponent
in the opposite time direction.  We re\"iterate that this symmetry can be violated, for short times, 
in response to inhomogeneities or to ``external perturbations''.\cite{b15}

By now many groups\cite{b1,b2,b3,b4,b5,b6} have illustrated the algebraic steps necessary to map the
``covariant'' exponents from one coordinate system to another.  A careless reader of some of this work
might well conclude (as we did) that ``covariant'' vectors and exponents are somehow coordinate-frame
independent.  A careful reader will instead note that because reference trajectories and nearby
satellite trajectories in one coordinate system can always be related to those in another, that the offset
vectors linking pairs of trajectories are likewise simply related so that the (different) exponents in both
frames can be computed.

It is not always emphasized that the exponents ( even the largest, which is ``covariant'' ) themselves
vary from frame to frame. For instance, in a useful and clarifying work, Posch\cite{b2,b9} selected a
spring-pendulum for his demonstration.  His two chosen frames were Cartesian and polar coordinates.
The constant-energy spring-pendulum dynamics can be described in either one of the three-dimensional
subspaces of the four-dimensional spaces in which the motion is described,  $(x,y,p_x,p_y )$ or
$(r,\theta,p_r,p_\theta)$ . Expressions linking the covariant exponents in these two frames (which
are different) are given in his paper.

Here we consider again the $(q,p,\zeta)$ oscillator, a {\it one}-dimensional rather than a two-dimensional
system, and described in a three-dimensional phase space.  The description can be carried out with
$\{ \ q,p,\zeta,\dot q,\dot p,\dot \zeta \ \}$ or with ``scaled variables''
$\{ \ Q,P,\zeta,\dot Q,\dot P,\dot \zeta \ \}$, where the two sets of variables are related by
the scaling $Q = 2q, P = (p/2)$ :
$$
\dot q = p \ ; \ \dot p = -q - \zeta p \ ; \ \dot \zeta = p^2 - T(q) \ ; \ T(q) = 1 + \epsilon \tanh (q) \ .
$$
$$
\dot Q = 4P \ ; \ \dot P = -(Q/4) - \zeta P \ ; \ \dot \zeta = 4P^2 - T(Q) \ ; \ T(Q) = 1 + \epsilon \tanh (Q/2) \ .
$$
Because the temperature depends upon the coordinate [ so that $T$ varies from
$( \ 1 - \epsilon \ )$ to $( \ 1 + \epsilon \ )$ ] , this model\cite{b10} is a generalized version of
the Nos\'e-Hoover oscillator described in Reference 12 .  The two sets of equations generate trajectories
which are {\it identical} if the coordinate and momentum axes are scaled because [ $Q = sq$ ] and
[ $P = (p/s)$ ] .  Here we compare $s = 1$ and $s = 2$ . The friction coefficient $\zeta $ ,  which directs
the squared momentum toward the local kinetic temperature $T(q)$ , is exactly the same function of time
in both the original unscaled and the scaled coordinate systems.  Thus the basic trajectories in $(qp\zeta)$
space and $(QP\zeta)$ space {\it are identical scale models of each other} apart from factors of two in the
directions associated with the length and momentum. Figure 3 shows the variation of the largest Lyapunov 
exponent with time along the relatively-simple limit cycle obtained when $\epsilon = 0.50$ .  Notice that
the local Lyapunov exponents $\lambda_1(t,q,p,\zeta,s)$ and $\lambda_1(t,Q,P,\zeta,s)$ {\it are} indeed
sensitive to the scale factor $s$ .

\begin{figure}
\vspace{1 cm}
\includegraphics[height=10cm,width=6cm,angle=-90]{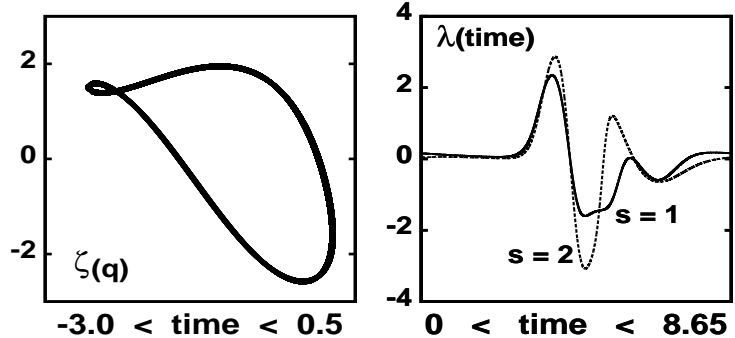}
\caption{
Local values of the largest Lyapunov exponents (right) for the limit cycle with $\epsilon = 0.50$ (left).  The 
time-averaged exponents are equal to $0.0$ .
}
\end{figure}

{\it This computation shows that the local exponents are quite different}.  Why is that? Here it is
because the stretch rates depend on the scale factor $s$.  The rates of stretching of pairs of ( infinitesimal )
tangent-space ``unit vectors'' parallel to $q = (s^{-1}Q)$ or parallel to $p = (s^{+1}P)$ are different :
$$
\dot \delta q = s^{-1}\dot \delta Q \ ; \ \dot \delta p = s^{+1} \delta P \ ,
$$
so that the corresponding local Lyapunov exponents in these two hypothetical cases would also vary
with $s$.  We exhibit this example here to emphasize the point that {\it even the local values of the
Lyapunov exponents depend on the chosen coordinate system}.  The global exponents for Hamiltonian systems
don't show this dependence.  In an email of 18 September 2013 Harald Posch showed that the global exponents
for a {\it doubly}-thermostated oscillator {\it do} depend on the scale factor $s$ but not on the norm.
Posch compared the exponents using both the usual $n=2$ norm and the unusual $n=3$ one :
$$
r^n = |dq|^n + |dp|^n + |d\zeta|^n + |d\xi|^n \ . 
$$

\section{Conclusion}
  Enthusiastic fans
of the MKS system of units cannot agree with the ardent fans of the cgs system when it comes to the local
exponents, either covariant or Gram-Schmidt.  Disinterested observers will note that one set of results
can be converted to the other, with the whole spectrum as well as its fluctuations dependent on the chosen
coordinate system.  The impression that ``covariant'' exponents are somehow uniquely special still seems
to us specious despite their norm-independence.

\section {Acknowledgments}
We thank Roger Samelson for the stimulating emails which led us careless readers to our current
understanding and reinvestigation of the dependence of the local covariant exponents on the chosen
coordinate system.  Likewise emails from Harald Posch were very useful.  They pointed out that scaling
transformations, such as the MKS-cgs distinction and the $s^{+1}$ and $s^{-1}$ combination, aren't usually
thought of as acceptable norms, although {\it no calculation at all} is possible without first choosing
{\it some} phase space.
We also appreciate our longstanding correspondences on this subject with Pavel Kuptsov
( see also his arXiv contributions ) and Franz Waldner as well as a useful remark by G\"unter Radons and
are happy to confess that at last we ``understand''.

\end{document}